\documentclass{article}

\usepackage{epsfig}
\usepackage{subfigure}
\usepackage{calc}
\usepackage{graphicx}

\usepackage{amsfonts}
\usepackage{amsmath}
\usepackage{amssymb}
\usepackage{latexsym}

\newcommand{\adots}{\mathinner{\mkern2mu\raise1pt\hbox{.}\mkern2mu%
\raise4pt\hbox{.}\mkern2mu\raise7pt\hbox{.}\mkern1mu}}

\begin{document}

\title{New modelling technique for aperiodic-sampling linear systems}
\date{}
\author{Amparo F\'{u}ster-Sabater and J.M. Guill\'en\\
{\small Instituto de Electr\'onica de Comunicaciones, C.S.I.C.}\\
{\small Serrano 144, 28006 Madrid, Spain} \\
{\small amparo@iec.csic.es}}

\maketitle

\begin{abstract}

A general input-output modelling technique for aperiodic-sampling
linear systems has been developed. The procedure describes the
dynamics of the system and includes the sequence of sampling
periods among the variables to be handled. Some restrictive
conditions on the sampling sequence are imposed in order to
guarantee the validity of the model. The particularization to the
periodic case represents an alternative to the classic methods of
discretization of continuous systems without using the
Z-transform. This kind of representation can be used largely for
identification and control purposes.


\end{abstract}

\section{Introduction}
\footnotetext{Work supported by Ministerio de Educaci\'{o}n y
Ciencia (Spain), Project TIC-0128.\\
International Journal of Control, Volume 45, Issue 3 March 1987, pages 951 - 968. \\
DOI: 10.1080/00207178708933780 } Aperiodic sampling is a very
interesting technique for improving the solution of several
problems in control and identification. In particular, aperiodic
sampling systems have been found useful in signal adaptation and
compensation \cite{Hsia74, Dormido80} and in optimal transmission
of measuring errors in problems involving the solutions of systems
of linear equations, such as observability \cite{Dormido79,
Troch73}, controllability \cite{Sen80} and identifiability
\cite{Sen79, Sen80}.

In all these cases it is essential to find a general model that
\begin{description}
\item (\textit{a}) describes the dynamics of the system;
\item (\textit{b}) is adapted to real experimentation
    conditions (external representation against internal-state
    representation);
\item (\textit{c}) includes the sequence of sampling periods
    among the variables to be handled.
    \end{description}

A well-known input-output formulation is used successfully for
linear systems sampled periodically. The scalar output of the
plant at an arbitrary instant is described by means of input and
output samples taken at previous instants. It seems natural to
consider the aperiodic case in the same way, except that the
finite difference equation coefficients, depending on the fixed
sampling period, would have to be replaced by multivariable
functions, depending on the aperiodic sampling sequence. Such a
formulation would satisfy the above conditions.

There must be restrictions on the aperiodic sampling sequence in
order that the model be valid, whereas in the periodic case any
sampling interval is valid.

The systems considered are characterized by their impulse response
(weighting function), which describes the output of the plant
through a convolution expression. Consequently, the properties and
results obtained are always given in terms of that function.

The paper is organized as follows. In \S 1 several basic
assumptions and general considerations, which will be assumed
through the work, are given. Section 2 is devoted to a modelling
technique for aperiodic-sampling linear systems. This also
includes the action of a sampler and zeroth-order hold preceding
the plant. Section 3 contains some restrictive conditions on the
sampling sequence. The problem of the selection of the aperiodic
sequence is strictly considered in geometric terms and the results
found in the literature are improved. In \S 4 a particularization
to the periodic case has been made. The proposed technique
represents an alternative to the classical methods of
discretization of continuous systems, in order to obtain the
discrete-model coefficients, without using the Z-transform.
Finally conclusions in \S 5 end the paper.

\section{Basic assumptions}
This discussion is restricted to:
\begin{description}
\item (i) linear time-invariant single-input/single-output
    differential systems of finite order $n$;
\item (ii) systems whose transfer function $G(S)$ is a
    strictly proper rational function, and whose impulse
    response $h(t)$ is therefore a particular solution of

\begin{equation}\label{eq:1}
h^{(n)}(t) + a_1h^{(n-1)}(t)+ \ldots + a_nh(t) = 0, \qquad t \geq 0
\end{equation}
an \textit{n}th-order homogeneous linear differential equation
with constant coefficients $(a_i \in \mathbb{R})$.
    \end{description}

$h(t)$ can be then be written as

\begin{equation}\label{eq:2}
h(t)= \sum\limits_{i=1}^{n} C_i \, \varphi_i(t), \qquad t \geq 0
\end{equation}

where $C_i \in \mathbb{C}$ are constant coefficients and
$\varphi_i; \mathbb{R} \rightarrow \mathbb{C} \;\; (i = 1,\ldots ,
n)$ is the fundamental system of solutions of eqn. (\ref{eq:1}).

We conclude this preliminary section with the following statement.
Let $(G_i)$ be a family of vector functions
\[G_i: \mathbb{R}^n \rightarrow  \mathbb{R}^n
\qquad (i=0,1, \ldots, n)
\]
\[G_i \in C^\infty(\mathbb{R}^n, \mathbb{R}^n)
\]

$C^\infty(\mathbb{R}^n, \mathbb{R}^n)$ being the set of infinitely
differentiable functions on $\mathbb{R}^n$. It then follows that
if there exist

\begin{description}
\item (\textit{a}) an integer $r \leq n$ such that the
    elements $(G_0(z),  \ldots$ $, G_r(z))$ are linearly
    independent for all $z \in \mathbb{R}^n$, and
\item (\textit{b}) an integer $k>r$ such that $G_k(z)$ depends
    linearly on $(G_0(z),  \ldots$ $, G_r(z))$, then there are
    functions
    \end{description}
\[f_0, f_1, \ldots , f_n \in C^\infty(\mathbb{R}^n, \mathbb{R})
\]
such that

\begin{equation}\label{eq:3}
\sum\limits_{i=0}^{n}f_{i}(z) \; G_i(z)=0  \qquad \forall z \in \mathbb{R}^n
\end{equation}

\section{Modelling technique}
\subsection{Generalities and key ideas of the methodology}
The class of linear time-invariant SISO systems is characterized
by the impulse response, which describes for zero initial
conditions the output of the plant through a convolution
expression

\begin{equation}\label{eq:4}
y_k=\sum\limits_{i=0}^{k}h(t_{k}-t_i) \, u_i
\end{equation}
where $y_k$ is the output of the plant at time $t_k$, $u_i$ is the
impulse input at time $t_i$, and $h(t)$ is the impulse response of
the plant. According to (\ref{eq:1}), the impulse response can
also be written in matrix form by means of the equivalent linear
system.

\begin{equation}\label{eq:5}
\dot{X}(t) = AX(t)
\end{equation}
where
\begin{equation}\label{eq:6}
X(t) = [\,h(t), \dot{h}(t), \ldots , h^{(n-1)}(t)\,]'
\end{equation}
the symbol $'$ denotes the transpose and $A$ is an $n \times n$
bottom-companion matrix.

\begin{equation}\label{eq:7}
A =
\left[%
\begin{array}{ccccc}
  0 & 1 & 0 & \ldots & 0 \\
  0 & 0 & 1 & \ldots & 0 \\
  \vdots & \vdots & \vdots & \, & \vdots \\
  -a_n & -a_{n-1} & -a_{n-2} & \ldots & -a_1 \\
\end{array}%
\right]
\end{equation}

The solution of the linear system
\begin{equation}\label{eq:8}
\dot{X}(t) = exp(At) X_0  \qquad (X_0 = X(0))
\end{equation}
is related to the impulse response through the expression

\begin{equation}\label{eq:9}
h(t) = c \; exp(At) X_0
\end{equation}
with
\begin{equation}\label{eq:10}
c = [\,1, 0, \ldots , 0\,]
\end{equation}

Note that the components of the vector $X_0$
\begin{equation}\label{eq:11}
X_0 = [\,h_1, h_2, \ldots , h_n\,]'
\end{equation}
correspond to the first \textit{n} Markov parameters

\begin{equation}\label{eq:12}
h_{i+1} = \frac{d^ih(t)}{dt^i}\bigg |_{t=0} \qquad (i = 0, 1,  \ldots, n-1)
\end{equation}

It should also be noted that the triad $(A, X_0, c)$ leads us naturally
to the observability canonical realization from the scalar impulse
response.

From these matrices $A, X_0, c$ we are going to define a family of
vectors functions

\[(G_0, G_1,  \ldots, G_n)
\]
\[G_i: \mathbb{R}^n \rightarrow  \mathbb{R}^n
\qquad (i=0,1, \ldots, n)
\]

given by
\begin{equation}\label{eq:13}
G_i(z_1, \ldots, z_n)= exp(A(z_1+ \ldots + z_i)) X_0  \qquad (i=1, \ldots, n)
\end{equation}

\begin{equation}\label{eq:14}
G_0(z_1, \ldots, z_n)= X_0
\end{equation}

with
\[ z = [\, z_1, z_2,  \ldots, z_n \,]' \in \mathbb{R}^n
\]

The functions $G_i$ and the impulse response are related by means
of the expression
\begin{equation}\label{eq:15}
h(z_1 + ... + z_i) =c \,G_i(z) \qquad z_1 + ... + z_i \geq 0
\end{equation}

Let us now consider the statement given in \S 1 for the kind of
function defined above.

From an analytical viewpoint, the functions $G_i$ belong to
$C^\infty(\mathbb{R}^n, \mathbb{R}^n)$, as compositions of
$C^\infty$ functions.

Let $I^n \subset \mathbb{R}^n$ be an open subset of $\mathbb{R}^n$
such that the vectors $(G_0(z),  \ldots$ $, G_{n-1}(z))$ are linearly
independent for all $z \in I^n$. In this case, it is easy to see
that for the new domain $I^n$ the conditions \textit{(a)} and
\textit{(b)} in the previous statement hold. In fact, condition
\textit{(a)} holds by definition of the subset $I^n$, and
condition \textit{(b)} holds by the dimensionality of the vector
$G_i$. Hence there will be functions
\[f_i(z) \in C^\infty(I^n, \mathbb{R}) \qquad (i=0, \ldots, n)
\]

such that

\begin{equation}\label{eq:16}
\sum\limits_{i=0}^{n}f_{n-i}(z) \; G_i(z)=0  \qquad \forall z \in I^n
\end{equation}

Rewriting (\ref{eq:16}) in matrix form, with
\[G_i(z) = [\,G_{1i}(z) \ldots  G_{ni}(z)\,]' \qquad (i=0, \ldots, n-1)
\]

we get
\begin{equation}\label{eq:17}
\left[
\begin{array}{ccc}
  G_{10}(z) & \ldots & G_{1,n-1}(z) \\
  \vdots & \, & \vdots \\
  G_{n0}(z) & \ldots & G_{n,n-1}(z) \\
\end{array}
\right]
\left[
\begin{array}{c}
  f_n(z) \\
  \vdots \\
  f_1(z) \\
\end{array}
\right] =
\left[
\begin{array}{c}
  G_{1n}(z) \\
  \vdots \\
  G_{nn}(z) \\
\end{array}
\right]  (-f_0(z))
\end{equation}

($f_0(z) = - 1$ for simplicity) and the functions $(f_1(z),
\ldots, f_n(z))$ can be obtained by solving a compatible system of
linear equations.

The general form of the functions $f_i(z)$ is

\begin{equation}\label{eq:18}
f_i(z) = \frac{Det[\,G_{0}(z) \ldots  G_{n}(z) \ldots G_{n-1}(z)\,]}{Det[\,G_{0}(z) \ldots  G_{n-1}(z)\,]}
\end{equation}

where the numerator is the determinant obtained from the matrix
\[[G_0(z), \ldots, G_{n-1}(z)]\] by replacing the \textit{i}th column by the column vector
$G_n(z)$.

At this point, we identify the components of $z$ with the elements
of the sampling period sequence

\begin{equation}\label{eq:19}
\left  \{ \begin{array}{r}
  z_n=t_k - t_{k-1} = T_k \\
  z_{n-1}=t_{k-1} -t_{k-2}=T_{k-1} \\
  \vdots \qquad \qquad  \;\;\;  \;\;\; \vdots \;\;\; \qquad \;\;\;  \vdots \;\;\;\\
  z_1=t_{k-n+1}-t_{k-n}=T_{k-n+1} \\
\end{array}  \right \}
\end{equation}

We multiply both sides of (\ref{eq:16}) by $c \, exp(Az^*)$, with
$z^*$ successively taking the values

\begin{equation}\label{eq:20}
z^* = - \sum\limits_{l=1}^{i} z_l \qquad (i=1, \ldots, n)
\end{equation}

and we get in each case
\begin{equation}\label{eq:21}
\left  \{ \begin{array}{rl}
  c \, exp(A(z_{i+1}+ \ldots + z_n)) X_0 = &  c \, \sum\limits_{l=1}^{n-i}f_{l}(z)\, exp(A(z_{i+1}+ \ldots + z_{n-l})) X_0 +\\
  \, & c \, \sum\limits_{l=0}^{i-1}f_{n-l}(z)\, exp(-A(z_{l+1}+ \ldots + z_{i})) X_0 \\
  \, & (i=1, \ldots, n)
\end{array}
\right \}
\end{equation}

We define
\begin{equation}\label{eq:22}
g_{n-i}(z)= c \, \sum\limits_{l=0}^{i-1}f_{n-l}(z)\, exp(A(z_{l+1}+ \ldots + z_{i})) X_0 \qquad (i=1, \ldots, n)
\end{equation}

From (\ref{eq:22}) the functions $g_{n-i}(z)$ are of class $C^\infty$
as compositions of $C^\infty$ functions.

It can be checked that the
functions $g_{n-i}(z)$ correspond to the first component of linear
combinations of vectors $G_i$ with negative argument.

If we write

\begin{equation*}
\left \{ \begin{array}{r}
        f_{i}^k =f_{i}(z) \\
        g_{j}^k =g_{j}(z)
      \end{array}
\right \}
\end{equation*}

at time $t_k$, we can condense the preceding
expressions into two sets of equations involving the functions
$f_{i}^k, g_{j}^k$ and $h(t)$:

\begin{equation}\label{eq:23}
\sum\limits_{i=0}^{n}f_{i}^k \,h(t_{k-i}-t{j})+g_{k-j}^k =0 \;\; (j=k, k-1, \ldots, k-n+1)
\end{equation}

\begin{equation}\label{eq:24}
\sum\limits_{i=0}^{n}f_{i}^k \,h(t_{k-i}-t_{j}) =0 \;\; (k \geq n ; \, j = k-n, \ldots, 0)
\end{equation}

Multiplying every equation by $u_j$ $(j = k, k - 1, \ldots ,
1,0)$ (impulse inputs at the sampling instants) respectively and
summing, we get
\begin{equation*}
f_1^k \,\sum\limits_{
l=0}^{k-1}u_{l}\, h(t_{k-1}-t_{l}) + \ldots +  f_n^k \,\sum\limits_{
l=0}^{k-n}u_{l} \,h(t_{k-n}-t_{l})
\end{equation*}
\begin{equation}\label{eq:25}
+\sum\limits_{j=0}^{n-1}g_{j}^k u_{k-j} = \sum\limits_{
l=0}^{k}u_{l} \,h(t_{k}-t_{l})
\end{equation}

and, according to (\ref{eq:4}), the preceding expression becomes

\begin{equation}\label{eq:26}
y_k = \sum\limits_{i=1}^{n}f_{i}^k y_{k-i} +  \sum\limits_{j=0}^{n-1}g_{j}^k u_{k-j}
\end{equation}

which is called the input-output model for linear
time-invariant aperiodic-sampling systems.

\subsection{Simplified form of the functions $f_i, g_j$}
Companion matrices are an important example
of what are called cyclic (or nonderogatory) matrices, which have
only one (normalized) eigenvector associated with each distinct
eigenvalue.
This means that

\begin{description}
\item (i) the Jordan canonical form is clearly simplified
    (there is only one Jordan block for each distinct
    eigenvalue);
\item (ii) the similarity transformation of the given matrix
    $A$ to the Jordan canonical form can be obtained in a
    standard way.
    \end{description}

Indeed,
\begin{equation}\label{eq:27}
A = BJB^{-1}
\end{equation}

where $J$ is the Jordan canonical form of the matrix $A$, and
$B$ is an invertible matrix of a well-known general form.

In this way, (\ref{eq:16}) becomes
\begin{equation}\label{eq:28}
\sum\limits_{i=0}^{n}f_{n-i}(z) \, exp(J\alpha_i) Y_0 = 0
\end{equation}

with
\begin{equation}\label{eq:29}
\alpha_i=z_1+ \ldots + z_i \qquad (i=1, \ldots, n)
\end{equation}

\begin{equation}\label{eq:30}
\alpha_0 = 0
\end{equation}

\begin{equation}\label{eq:31}
Y_0 = B^{-1} X_0
\end{equation}

Factorizing $det[exp (J \alpha_0) Y_0, \ldots, (J \alpha_{n-1})
Y_0]$ by means of Laplace's expansion by minors and cancelling
common factors in the numerator and denominator of (\ref{eq:18}),
the functions $f_i^k$ can be written as

\begin{equation}\label{eq:32}
f_i^k = \frac{\Delta_i}{\Delta} \qquad (i=1, \ldots, n)
\end{equation}

where

\begin{equation}\label{eq:33}
\left \{ \begin{array}{cc}
  \Delta = | \varphi_l (\alpha_j)| &  (j=0, 1, \ldots, n-1) \\
  \Delta_i = | \varphi_l (\alpha_j)| &  (j=0, \ldots, n, \ldots, n-1) \\
  \, & (l=1, \ldots, n)
\end{array} \right \}
\end{equation}

The linear independence of the vectors $G_i$
implies the non-nullity of the determinant $\Delta$. We can check that
the functions $f_i$ depend exclusively on the poles of the transfer
function and on the sampling sequence. The result could be
expected and agrees for the periodic case with the direct
correspondence
\[\lambda_i \rightarrow  exp(\lambda_i T_0)
\]

between the poles of the continuous and pulse transfer function.

The functions $g_j^k$ are given
by

\begin{equation}\label{eq:34}
g_{k-j}^k = -\, \sum\limits_{i=0}^{n}f_{i}^k \, h(t_{k-i}-t_{j}) \qquad (j=k, k-1, \ldots, k-n+1)
\end{equation}

and depend on the poles and zeros of the transfer function, since
they are obtained as a linear combination of the impulse response
for specific arguments.

\subsection{General formulation for systems with
zeroth-order hold}
If the sampler at the input is followed by a
zeroth-order hold the input-output model is modified in the
following way:
\begin{equation}\label{eq:35}
y_k=\sum\limits_{i=0}^{k}h(t_{k}-t_i) \, x_i
\end{equation}

with

\begin{equation}\label{eq:36}
x_i = u_i - u_{i-1}
\end{equation}

\begin{equation}\label{eq:37}
h(t)= \mathcal{L}^{-1}  \left[
\begin{array}{c}
\frac{G(s)}{s}
\end{array}\right]
\end{equation}

As $x_i$ represents the difference between two consecutive values of the
input signal, the expressions (\ref{eq:23}) and (\ref{eq:24}) become
\begin{equation}\label{eq:38}
\left \{ \begin{array}{cc}
  \sum\limits_{i=0}^{n}f_{i}^k[\, h(t_{k-i}-t_j) - h(t_{k-i}-t_{j+1})\,] + g_{k-j}^k=0 &  (j=k, k-1, \ldots, k-n)  \\
  \sum\limits_{i=0}^{n}f_{i}^k[\, h(t_{k-i}-t_j) - h(t_{k-i}-t_{j+1})\,] =0 &  (j=k-n-1,  \ldots, 0)  \\
  \, & (k>n)
\end{array} \right \}
\end{equation}

Briefly, the presence of a zeroth-order hold
implies that
\begin{description}
\item (\textit{a}) the functions $f_i$ are the same as before;
\item (\textit{b}) the functions $g_j$ are modified: the
    number functions $g_j$ is increased by one (so that now
    $0\leq j \leq n$), and their general expression is similar
    to (\ref{eq:34}) but with
\[h(t_{k-i}-t_j) \rightarrow h(t_{k-i}-t_j)- h(t_{k-i}-t_{j+1})
\]
for $h(.)$ defined in (\ref{eq:37}).
\end{description}

The results obtained were
to be expected, since the functions $f_i$ depend only on the poles of
the transfer function, which reflect the internal coupling in the
system and its autonomous behaviour. However, the functions $g_j$
reflect the internal plant coupling to the input signal, which has
been affected by the presence of the zeroth-order hold.

\subsection{Main results}
We recall briefly the main results we have obtained.
\begin{description}
\item (\textit{a}) We have shown that the functions $f_i$,
    $g_j$ are infinitely differentiable.
\item (\textit{b}) We have obtained a general and systematic
    formulation for every function  $f_i$, $g_j$ in contrast
    with the procedure found in \cite{Mellado70}, where all
    these functions are computed globally.
\item (\textit{c}) We have developed a procedure to impose
    restrictive conditions on the sampling sequence in order
    to guarantee the linear independence of the vectors
    $(G_0(z), G_1 (z), \ldots , G_{n - 1}(z))$.
\end{description}
In the next section we shall determine the
set of vectors $I^n \subset \mathbb{R}^n$ whose elements $z$ satisfy the above
condition.

\section{Choice of the sampling-period sequence. A geometric
interpretation} The problem of the choice of the sampling sequence
has been treated in \cite{Troch73} analytically. Making use of the
concept of a Chebyshev system, some intervals of the real line are
selected in which the sampling instants can be chosen freely. In
this way, the non-nullity of the determinant $\Delta$ is
guaranteed, and consequently the difference equation for the
aperiodic case can be obtained (via (\ref{eq:26}), (\ref{eq:32})
and (\ref{eq:34})).

In the present paper the same problem is considered
geometrically. The choice of sampling-period sequence is directly
related to the properties of certain vectors in the space $\mathbb{R}^n$. In
this way, the intervals found by Troch \cite{Troch73} are largely increased for
low-order models and some general considerations for higher-order
models are made.

\subsection{Second-order model (n = 2)}
We are going to consider a second-order model with a pair of complex eigenvalues
\begin{equation}\label{eq:39}
a + jb \in \mathbb{C} \qquad (b > 0)
\end{equation}

The problem depends on an adequate choice
of the sampling periods in such a way that the linear independence
of the vectors
\begin{equation}\label{eq:40}
[Y_0, Y_1] = [exp (J\alpha_0)\,z_0, exp (J\alpha_1)\,z_0]
\end{equation}
will
be preserved, where
\begin{equation}\label{eq:41}
\alpha_0 = 0
\end{equation}
\begin{equation}\label{eq:42}
\alpha_1 = t_{k-1} - t_{k-2}= T_{k-1}
\end{equation}
$J_r$  is the real
canonical form of the matrix $A$:

\begin{equation}\label{eq:43}
J_r = \left[
    \begin{array}{cc}
      a & -b \\
      b & a \\
    \end{array}
  \right]
\qquad (b > 0)
\end{equation}
and $T$ is the (invertible) matrix governing the change of basis:
\begin{equation}\label{eq:44}
z_0 = T^{-1}X_0
\end{equation}

As we are in $\mathbb{R}^2$, the geometric
interpretation is very simple. The generic operator $exp (J_r\alpha)$
applied to the vector $z_0$ can be viewed as follows. It is a
counterclockwise rotation through $b\alpha$ radians, followed by a
stretching (or shrinking) of the length of $z_0$ by a factor $exp (a\alpha)$ \cite{Hirsch74}.

From this interpretation $Y_0$ and $Y_1$, will be linearly independent if and only if

\begin{equation}\label{eq:45}
b\alpha_1 = b\,T_{k-1}\neq {\dot{\pi}}
\end{equation}
where ${\dot{\pi}}$ denotes an integral multiple of $\pi$. Otherwise the vectors will be
colinear.

Comparing these results to those of Troch \cite{Troch73} we have
the following. According to \cite{Troch73}, given $t_{k-2}$, $t_{k-1}$ can be
fixed so that
\begin{equation}\label{eq:46}
t_{k-1} \in (t_{k-2}, t_{k-2} + \frac{\pi}{b})
\end{equation}

According to (\ref{eq:45}), given $t_{k-2}$, $t_{k-1}$ can be
fixed so that
\begin{equation}\label{eq:47}
t_{k-1} \in I = (t_{k-2}, \infty ) - (t_{k-2} + \frac{ {\dot{\pi}} }{b})
\end{equation}

Only point values will be rejected.

It can also be checked that such a formulation is concerned with
differences between sampling instants but not with the actual
position of such instants on the real axis. This agrees perfectly
with the kind of time-invariant systems we are dealing with.

If the condition (\ref{eq:45}) is violated, the sampling process
resonates with the system dynamics and the formulation then
obtained no longer affords a faithful representation of the
system.

We can also see that, for the second-order model, (\ref{eq:45}) is a
necessary and sufficient condition to guarantee the linear
independence of the vectors $Y_i$ whereas the condition given by
Troch is only sufficient.

\subsection{Third-order model (n = 3)}
We are going to consider a third-order model with a real pole and a
complex pair

\begin{equation}\label{eq:48}
\lambda \in \mathbb{R}, \;\; a + jb \in \mathbb{C} \;\, (b > 0)
\end{equation}

The point is to choose the
sampling instants in the right way; that is, such that the vectors

\begin{equation}\label{eq:49}
[Y_0, Y_1, Y_2] = [exp (J_r\alpha_0)\,z_0, exp (J_r\alpha_1)\,z_0, exp (J_r\alpha_2)\,z_0]
\end{equation}
are linearly independent, where
\begin{equation}\label{eq:50}
\alpha_0 = 0
\end{equation}
\begin{equation}\label{eq:51}
\alpha_1 = t_{k-2} - t_{k-3}= T_{k-2}
\end{equation}
\begin{equation}\label{eq:52}
\alpha_2 = t_{k-1} - t_{k-3}= T_{k-1} + T_{k-2}
\end{equation}

\begin{equation}\label{eq:53}
J_r = \left[
      \begin{array}{ccc}
        a & -b & 0 \\
        b & a & 0 \\
        0 & 0 & \lambda \\
      \end{array}
    \right]
\qquad (b>0)
\end{equation}

with $z_0, T$ defined as before.

The problem is treated in three-dimensional space, and the
geometric structure can be described as follows. The generic vector $Y(\alpha)$ is written as
\begin{equation}\label{eq:54}
Y(\alpha) = [exp (a \alpha)\, cos(b \alpha), exp (a \alpha) \, sin(b \alpha), exp (\lambda \alpha)]'
\end{equation}



It is therefore, a
spiral on the surface of revolution
\begin{equation}\label{eq:55}
 z = (x^2+ y^2)^{\lambda/2a}
\end{equation}

whose form is determined by the real part of the system
eigenvalues.

The vectors $Y_0, Y_1, Y_2$ have their origin at the point
$(0, 0, 0)$ and their ends at the points $Y_(\alpha_0), Y_(\alpha_1), Y_(\alpha_2)$ of the
parametric curve.

Heuristically we can imagine the same vectors as
before ($n = 2$) pointing upwards from the $(X, Y)$ plane as they have
a third component on the $Z$ axis.

From this geometric
interpretation we can study some interesting cases of linear
dependence.


\begin{description}
\item (\textit{a}) If the sampling-period sequence is chosen
    in such a way that
\begin{equation}\label{eq:56}
 b \alpha_i = {\dot{\pi}} \qquad (i = 0, 1, 2)
\end{equation}
according to Fig. 1, the vectors $Y_0, Y_1, Y_2$ will be
coplanar (in a plane containing the $Z$ axis) and therefore
they will be linearly dependent since every vector can be
written as a linear combination of the other two. In physical
terms, this can be regarded as a resonance of the sampling
sequence with a pair of complex eigenvalues. This situation
can be generalized to higher-order models if the equation
(\ref{eq:56}) holds for the imaginary part of any pair of
complex eigenvalues.
\item (\textit{b}) If the real eigenvalue and the real part of
    the complex pair are equal then the revolution surface is
    a cone and there will be linear dependence if at least two
    vectors are on the same generator (see Fig. 2).
    Analytically,
\begin{equation}\label{eq:57}
 b (\alpha_j - \alpha_i) = 2{\dot{\pi}}
\end{equation}

for any $i,j$ such that $0\leq i < j \leq 2$.

Physically, this situation can be regarded as a
resonance of all the eigenvalues with the time interval between
two sampling instants, not necessarily consecutive.

This situation can be generalized to higher-order models if the
eigenvalues are such that
\begin{description}
\item (i) the real parts are the same, and
\item (ii) all the imaginary parts $b_j \; (j = 1, \ldots,
    q)$ satisfy (\ref{eq:57}).
\end{description}

\item (\textit{c}) This is the general case where there is
    linear dependence because the vectors $Y_i$ are contained
    in an arbitrary plane, which obliquely intersects the
    surface of revolution passing through the origin.
\end{description}


The intersection of the oblique plane and the surface of
revolution is a closed curve $\Gamma$. Therefore the spiral will
intersect such a curve in only two points for each rotation of
$2\pi$ radians. From this interpretation, different geometric
structures can be obtained, depending on the parameter values. In
fact, we have the following possibilities:
  \[ \begin{array}{cc}

  \lambda > 0 &
    \left \{ \begin{array}{c}
    a>0 \\
    a<0  \end{array} \right \} \\
     \, & \, \\
   \lambda < 0 &
    \left \{ \begin{array}{c}
    a>0 \\
    a<0  \end{array} \right \}

\end{array}  \]

Now $\lambda > 0 \, (\lambda < 0)$ implies that the spiral goes up
(down) as far as $\alpha \rightarrow \infty$; and $ a > 0 \, (a <
0)$ implies that the surface of revolution expands (shrinks) when
we move up the $Z$ axis. \vspace*{0.2cm}

\textit{Case 1:} $\lambda >0, a>0$

From examination of $P_0$ and $P_1$ the normalized projections of
$Y_0$ and $Y_1$ on the $(X, Y)$ plane in Fig. 4, it is clear that
if $Y_2$ is contained in the oblique plane then its normalized
projection on the $(X, Y)$ plane is situated on the $arc P_0 P_1 =
b\alpha_1$ (here $P_0$ and $P_1$ denote the ends of the vectors
$P_0$ and $P_1$ respectively). Therefore a sufficient condition
that guarantees the linear independence of the three vectors is
that the rotation $b \alpha_2$ is not part of such an arc.

In this way, $P_0 P_1$ and the corresponding multiples of $2\pi$
are the forbidden intervals, and consequently $P_1 P_0$ and the
corresponding multiples of $2\pi$ are the allowed ones.

It has to be pointed out that inside the forbidden
intervals there would only be two intersection points of the
spiral with the curve $\Gamma$ for each rotation through angle $2\pi$.
However, we reject the whole interval since we intend to find
sampling intervals in which the linear independence of the vectors
$Y_i$ is automatically guaranteed without analytical computations.

The most favourable case corresponds to a rotation $P_0 P_1$ as
short as possible, because this would increase the length of the
allowed interval. Reciprocally, the most disfavourable case
corresponds to a rotation $P_0 P_1$, with angle close to $\pi$.

For values of $\alpha$ such that

\begin{equation}\label{eq:58}
 |Y(\alpha)| > |Y(\alpha^*)|
\end{equation}

where $Y(\alpha^*)$ is the vector of maximum
modulus with its origin at the point $(0, 0, 0)$ and its end on the
curve $\Gamma$, the choice of the third vector is completely arbitrary,
since $Y_2$ can never be on the oblique plane. Therefore the number
of forbidden intervals is finite and depends on the previous
relation.

For the case $\lambda > 0, a < 0$ the allowed intervals are the
same as before.

\vspace*{0.2cm}

\textit{Case 2:} $\lambda < 0, a < 0$


We proceed in the same way as before. Figure 5 shows the
projections of $Y_0, Y_1$ and the curve $\Gamma$ on the $(X, Y)$
plane. The diameter $R_1R_0$ is parallel to the chord joining the
end of the normalized projections $P_0$ and $P_1$. For each
rotation through angle $2\pi$, the spiral will intersect the curve
$\Gamma$ in only one point at $P_1 R_1$, and the same will happen
at $R_0 P_0$. Therefore a sufficient condition to guarantee the
linear independence of the three vectors in that the rotation
$b\alpha_2$ is not part of such an arc. Consequently, $P_0 P_1$
and $R_1 R_0$ and the corresponding multiples of $2\pi$ radians
would be the allowed intervals.

It should be pointed out that the arc $P_1 R_1$ is allowed
only for the first rotation. The most favourable case corresponds
to a rotation $P_0 P_1 = b \alpha_1$, with angle as close to $\pi$ as possible.
Conversely, the most unfavourable case corresponds to a rotation
with angle as small as possible.

In this situation the number of forbidden intervals is not finite,
since there are always two intersection points for each rotation
of angle $2 \pi$.

For the case
$\lambda < 0, a> 0$ the allowed intervals are the same as before.

Comparing these results with those of \cite{Troch73}, we can see
that for the latter the allowed interval, in rotational form, can
be expressed as

\begin{equation}\label{eq:59}
 b\alpha_2 < \pi
\end{equation}

which, according to the
geometric interpretation, is clearly a shorter interval. In both
cases the conditions are sufficient.

The only interesting cases for second- and
third-order models are those considered in $\S\S$ 4.1 and 4.2.

For systems without complex eigenvalues, the linear independence of
the vectors $Y_i$ is automatically guaranteed for any arbitrary
choice of the sampling instants \cite{Troch73}.


\subsection{General considerations for higher-order models}
From the fourth-order upwards, the situation becomes more and more complex because we do
not have the geometrical insight given by the plane or
three-dimensional space. However, from the basic structures
developed for $\mathbb{R}^2$ and $\mathbb{R}^3$, we can make some general
considerations that simplify the choice of sampling instants in
higher-order models.

In order to clarify these ideas, we are going
to consider a 4th-order model ($n = 4$) with one pair of complex
eigenvalues and one pair of real eigenvalues:
\begin{equation}\label{eq:60}
\lambda_1, \lambda_2 \in \mathbb{R}, \;\; a + jb \in \mathbb{C}
\end{equation}

The parametric curve is
\begin{equation}\label{eq:61}
Y(\alpha) = [exp (a \alpha) \,cos(b \alpha), exp (a \alpha) \,sin(b \alpha), exp (\lambda_1 \alpha), exp (\lambda_2 \alpha)]'
\end{equation}

which can be decomposed into
\begin{equation}\label{eq:62}
c_1(\alpha) = [exp (a \alpha) \,cos(b \alpha), exp (a \alpha) \,sin(b \alpha), exp (\lambda_1 \alpha)]
\end{equation}

\begin{equation}\label{eq:63}
c_2(\alpha) = [exp (a \alpha) \,cos(b \alpha), exp (a \alpha) \,sin(b \alpha), exp (\lambda_2 \alpha)]
\end{equation}

and we can study the evolution of each curve separately in the same way as before.

The
normalized projections of the vectors
\begin{equation*}
(c_1(\alpha_i)), (c_2(\alpha_i)) \qquad (i= 0, \ldots, 3)
\end{equation*}

on the $(X, Y)$ plane will coincide, since the only element
varying is the third component and not the angle of rotation.

Take $(\alpha_0, \alpha_1,\alpha_2,\alpha_3)$ such that each
subset of three elements in either of the sets $(c_1(\alpha_i))$
$(c_2(\alpha_i))$ is linearly independent. Then we have that each
subset of three elements in $Y_i$ $(i = 0, \ldots, 3)$ is linearly
independent. Therefore it suffices to check the linear dependence
of any one vector with respect to the other three.

In analytical terms, this means that
\begin{equation}\label{eq:64}
M_1 \,c_1(\alpha_j)= M_2 \,c_2(\alpha_j)
\end{equation}

where $M_1$ and $M_2$ are $3 \times 3$ matrices of general form

\begin{equation}\label{eq:65}
\left \{ \begin{array}{c}
  M_{1}= (c_1(\alpha_i))  \\
  M_{2}= (c_2(\alpha_i))
\end{array} \right \} \qquad (i=0,  \ldots, 3), \; i \neq j
\end{equation}

If (\ref{eq:64}) does not hold
then the linear independence of the vectors $Y_i$ $(i = 0, \ldots, 3)$,
obtained through the manipulation of 3-dimensional vectors, is
guaranteed. Therefore we have been able to reduce the problem
dimension by one.

The method can be generalized to arbitrary order
$n$, although the problem becomes more and more complicated as the
order of the model is increased.

If the multiplicity of the poles
is greater than 1, the third component of the vectors $c_i(\alpha)$ is a
polynomial expression with trigonometric functions, which will
give a more complicated curve.

If there is more than one pair of complex eigenvalues, that with
the greatest imaginary part (maximum eigenfrequency) will be the
main pair; since it determines the rotation with greater
angle.

When the dimension of the model is very large, it is more
convenient to use the shorter intervals found in the literature or
to reduce the order of the model and then apply the techniques
developed to the reduced model.

\section{Particularization to the periodic case}
The aperiodic model developed in previous
sections is also valid for the periodic case. In this situation,
the general expressions and the geometric interpretation can be
simplified.

\subsection{Periodic model}
Once the sampling period $T_0$ has been
fixed, the multivariable functions $f_i, g_j$ are reduced to
constant coefficients $a_i, b_j$.

It can be demonstrated by induction that for the one-variable case
the coefficients $a_i$ can be written as

\begin{equation}\label{eq:66}
a_i= (-1)^{i-1} \, \Sigma \phi_{j1} \ldots \phi_{ji}
\end{equation}

where the sum is taken for $1 \leq jp \leq n$ and $1 \leq p \leq i
\; (i = 1, ldots, n)$
\begin{equation}\label{eq:67}
\phi_{ji}(T_0)= exp(\lambda_i T_0)
\end{equation}

($\lambda_i$ is an eigenvalue of the continuous system). If the
multiplicity of $\lambda_i$ is $m_i > 1$ then there are $m_i$
functions $\phi_{i}$ which are all equal. The $a_i$ correspond to
coefficients with opposite sign of the polynomial $A(z)$ in the
$Z$-transfer function
\begin{equation}\label{eq:68}
G(z)= \frac{B(z)}{A(z)}
\end{equation}

Indeed,
\begin{equation}\label{eq:69}
A(z) = 1+ a'_1 z^{-1} + \ldots + a'_n z^{-n}
\end{equation}

with
\begin{equation}\label{eq:70}
a'_i = -a_i
\end{equation}

It is well known that between the coefficients and the roots of a
polynomial there is a variational relation like that given by
(\ref{eq:66}).

According to (\ref{eq:34}), the coefficients $b_j$ are
\begin{equation}\label{eq:71}
b_{k-j}= - \sum\limits_{i=0}^{n}a_{i} \,h(t_{k-i}-t_{j}) \;\; (j = k, k-1, \ldots, k-n+1)
\end{equation}

with $a_0= -1,\, k \geq n$. Equations (\ref{eq:66}) and
(\ref{eq:71}) represent an alternative approach to that presented
by the $Z$-transform in order to compute discrete-model
coefficients from the weighting function.

The formulation corresponding to the case with zeroth-order hold
can be also transcribed to the periodic version by means of
(\ref{eq:38}).

\subsection{Influence of sampling period on the
discrete model parameters} To discuss the effect of sampling
period on the absolute values of parameters we give the following
example.

We consider the continuous transfer function
\begin{equation}\label{eq:72}
G(s)= \frac{K}{(1+T_1s) \, (1+T_2s)\, (1+T_3s)}
\end{equation}

with zeroth-order hold. The external representation is
\begin{equation}\label{eq:73}
y_k = \sum\limits_{i=1}^{3}a_{i} y_{k-i} +  \sum\limits_{j=0}^{3}b_{j}^k u_{k-j}
\end{equation}

where $a_i$ and $b_j$ can be obtained according to (\ref{eq:66})
and (\ref{eq:71}) with
\begin{equation}\label{eq:74}
\phi_i(T_0) = exp(-\frac{1}{T_i}T_0) \qquad (i=1, 2, 3)
\end{equation}

For the following values of the continuous model parameters
\begin{equation*}
K = 1, \; T_1 = 10 s, \; T_2 = 7.5 s, \; T_3 = 5 s
\end{equation*}

the values of the discrete-model coefficients are shown in Table
 (\ref{table:headings1}) for different sampling periods $T_0$.

\setlength{\tabcolsep}{4pt}
\begin{table}
\begin{center}
\caption{Values of the parameters for different sampling periods}
\label{table:headings1}
\begin{tabular}{cccccccccc}
\noalign{\smallskip} \hline
$T_0$ & $2$ & $4$ & $6$ & $8$ & $10$ & $12$ \\
\hline \noalign{\smallskip}
$b_1$ & $0.0026$ & $0.0186$ & $0.0510$ & $0.0989$ & $0.1586$ & $0.2260$ \\
$b_2$ & $0.0092$ & $0.0486$ & $0.1086$ & $0.1718$ & $0.2257$ & $0.2643$ \\
$b_3$ & $0.0018$ & $0.0078$ & $0.0139$ & $0.0174$ & $0.0181$ & $0.0167$ \\
$a_1$ & $2.2549$ & $1.7063$ & $1.2993$ & $0.9953$ & $0.7668$ & $0.5938$ \\
$a_2$ & $-1.689$ & $-0.958$ & $-0.547$ & $-0.314$ & $-0.182$ & $-0.106$ \\
$a_3$ & $0.4203$ & $0.1767$ & $0.0742$ & $0.0312$ & $0.0131$ & $0.0055$ \\
$\Sigma b_i= 1 + \Sigma a_i$ & $0.0139$ & $0.0750$ & $0.1736$ & $0.2882$ & $0.4025$ & $0.5071$ \\
\hline
\end{tabular}
\end{center}
\end{table}

These results agree with those obtained by Isermann
\cite{Isermann81}. Because of the sign of $T_i$ and the general
form of $a_i,\, b_j$, the magnitudes of the $a_i$ parameters
decrease (in absolute value) and those of $b_j$ increase with
increasing sampling period $T_0$.

If $T_i$ had the opposite sign the parameter behaviour would be
entirely different.

For a small sampling period $T_0 = 1s$

\begin{equation*}
b_i\ll |a_i|, \;  \Sigma b_i \ll |a_i|
\end{equation*}

This is why small errors in the estimated parameters can have a
significant influence on the input-output behaviour of the model.
Indeed, $\Sigma b_i$ depends on the 4th or 5th place of $b_i$
after the decimal point.

If the sampling period is chosen too small, ill-conditioned
matrices result, which leads to numerical problems. On the other
hand, if the sampling period is chosen too large then the
dynamical behaviour is described inexactly. For $T_0 = 10$ s the
model is practically reduced to second order because

\begin{equation*}
a_3\ll 1 + \Sigma|a_i|, \;  b_3\ll \Sigma b_i
\end{equation*}
and for even greater sampling periods we get a first-order model.

A proper choice of sampling interval in most cases is not
critical, because the range between too-small and too-large values
is relatively wide \cite{Isermann81}.

\subsection{Geometric particularization}
The geometric interpretation of \S 3 can be particularized to the
periodic case. We are going to consider again the third order
model because it is the most significant. For the cases previously
presented, we get the following results.
\begin{description}
\item (\textit{a}) The vectors $Y_i$ are coplanar if and only
    if
\begin{equation}\label{eq:75}
bT_0 = {\dot{\pi}}
\end{equation}
\item (\textit{b}) This is included in the previous case.
\item (\textit{c}) This situation never occurs, since
    different rotations of the vectors $Y_i$ would imply
    different sampling periods, which is not possible in the
    periodic case. \end{description}

We can see that the transition from the aperiodic case to the
periodic case is a simple particularization. In the opposite
sense, the procedure is much more complicated, because there are
many aperiodic situations that have no equivalent in the periodic
version.

In analytical terms, the coefficients $a_i$, and consequently the
$b_j$, are perfectly defined for all sampling periods $T_0$,
according to (\ref{eq:66}) and (\ref{eq:71}). We recall that the
$Z$-transform imposes no constraints on the choice of $T_0$.
Conversely, in the aperiodic formulation there are some sampling
sequences that are forbidden for the model developed. This
confirms the complexity of the aperiodic case compared with the
periodic one.

\subsection{Coefficients of the discrete model for systems with dead time}
We are going to consider the same system with zeroth-order hold as
before, but also with dead time $T_d$. The convolution expression
that reflects this situation is

\begin{equation}\label{eq:76}
y_k=\sum\limits_{i=0}^{k}h(t_{k}-t_i-T_d) \, x_i
\end{equation}

with $x_i$ and $h(t)$ defined as before. For

\begin{equation}\label{eq:77}
(p-1)T_0 < T_d \leq p T_0
\end{equation}

with $p \in Z$, the general input-output model can be written as

\begin{equation}\label{eq:78}
y_k = \sum\limits_{i=1}^{n}a_{i} y_{k-i} +  \sum\limits_{j=0}^{n}b_{j} u_{k-j-p}
\end{equation}

In order to determine the coefficients $a_i$ and $b_j$, the
procedure is similar to that previously developed. Indeed, now we
have the expressions

\begin{equation}\label{eq:79}
  \sum\limits_{i=0}^{n}a_{i}[\, h(t_{k-i}-t_j-T_d) - h(t_{k-i}-t_{j+1}-T_d)\,] + b_{k-j}=0
\end{equation}
with $(j=k-p, k-p-1, \ldots, k-p-n)$.
\begin{equation}\label{eq:80}
  \sum\limits_{i=0}^{n}a_{i}[\, h(t_{k-i}-t_j-T_d) - h(t_{k-i}-t_{j+1}-T_d)\,] =0 \; (j=k-p-n-1, \ldots, 0)
\end{equation}

and we see that for $j = k, k - 1, \ldots,k - p + 1$
\begin{equation}\label{eq:81}
 h(t_k-t_j-T_d)=0
\end{equation}

There are two different cases.
\begin{description}
\item (i) $T_d = pT_0$, which implies that the coefficients
    $a_i$ are not modified by the dead time; the coefficients
    $b_j$ are not modified either, except that these
    coefficients multiply the impulse inputs $u_{k-j-p}\;\; (j
    =0,1, \ldots , n)$, rather than the inputs $u_{k-j}\;\; (j
    =0,1, \ldots , n)$ for the case without dead time.
    \item (ii) $(p - 1)T_0 < T_d < pT_0$, which implies that
        the coefficients $a_i$ are not modified by the dead
        time; the coefficients $b_j$ are now functions of
        $T_d$ whose general expression is given by
        (\ref{eq:79}) and which multiply the inputs
        $u_{k-j-p}\;\; (j =0,1, \ldots , n)$.
        \end{description}


\section{Conclusions}
A general aperiodic model for linear time-invariant SISO systems
has been developed. The model also covers more general cases such
as systems with zeroth-order hold and dead time. The formulation
considered stresses the importance of the sampling period against
other system parameters. In this way, such systems have an
additional element for analysis and manipulation. The
periodic-sampling case appears as a simple particularization of
the general procedure. The results obtained are simplified and the
use of tables of $Z$-transforms are avoided. For every sampled
system, in a periodic or aperiodic way, there will always be
sampling-period sequences more or less adequate according to the
general characteristics of the process under study. In the
aperiodic case, there will also be some restrictive conditions on
these sequences, although it has also been possible to give
strategies for some special cases.


\end{document}